\documentclass[12pt]{elsart}
\usepackage{epsfig,amssymb}
\usepackage{refmerge}
\begin{document}
\newcommand{\beqn}{\begin{equation}}
\newcommand{\eeqn}{\end{equation}}
\newcommand{\ee}{e^+e^-}
\newcommand{\et}{\eta^1_1}
\begin{frontmatter}
\date{}

\title{\Large \bf Pion Form Factor Phase, $\pi\pi$ Elasticity and
New $\ee$ Data}

\author[BINP]{S.~Eidelman} and
\author[IFJ]{L.~{\L}ukaszuk}

\address[BINP]{Budker Institute of Nuclear Physics, Acad. Lavrentyev 11,
Novosibirsk, 630090, Russia}
\address[IFJ]{Andrzej So{\l}tan Institute for Nuclear Studies,
Ho\.{z}a 69, PL-00-681,  Warsaw, Poland }

\begin{abstract}
New precise data on the low energy $\ee$ annihilation into hadrons 
from Novosibirsk are used to obtain bounds on the elasticity
parameter and the difference between
the phase of the pion form factor and that of the $\pi\pi$ scattering.
\end{abstract}
\end{frontmatter}
\maketitle

Pion form factor and its relation to $\pi\pi$ scattering have been  
extensively studied for many years (see~\cite{roy} and references
therein).
Although the form factor phase naturally appears in any 
model of the pion form factor~\cite{ph1,ph2,ph3,ph4,ph5}, 
it is well known that only 
the  absolute value of the form factor can be usually measured while 
information on the phase can be gained from sophisticated interference 
experiments.  
However, as shown long ago, there is an interesting possibility to obtain 
bounds on the elasticity parameter of the P-wave $\pi\pi$ scattering,
$\eta_1$, and the
difference between the phase of the pion form factor $\psi$ and 
that of the $\pi\pi$ scattering $\delta_1$ in a model-independent 
way under very general assumptions~\cite{luk}. Namely, the 
following inequality has been obtained there\footnote{A factor
$|F|^2$ was unfortunately omitted on l.h.s. of formula (5a) in 
Ref.~\cite{luk}. All other relations in Ref.~\cite{luk} are correct.}:
\beqn
(\frac {1-\eta_1} {2})^2 + \eta_1 \sin^2{(\psi - \delta_1)} \leqslant
\frac {1-\eta_1^2} {4} \cdot r,~~~~0 \leqslant \eta_1 \leqslant 1
\label{gen1}
\eeqn
or, equivalently,
\beqn
|a(\eta_1, \psi - \delta_1)|^2 \leqslant
\frac {1-\eta_1^2} {4} \cdot r
\label{gen2}
\eeqn
with $a(\eta,\alpha) \equiv (\eta \exp{2i\alpha}-1)/2i$.
Here $r$ is the following ratio:
\beqn
r=\frac {\sigma^{\rm I=1}_{\rm e^+e^-}} 
{\sigma_{\rm e^+e^- \to \pi^+\pi^-}} - 1=
  \frac {\sigma^{\rm I=1}_{\rm  non-2\pi} +
        \sigma_{\rm e^+e^- \to \pi^+\pi^-}}
  {\sigma_{\rm e^+e^- \to \pi^+\pi^-}} - 1=
  \frac {\sigma^{\rm I=1}_{\rm  non-2\pi}}
 {\sigma_{\rm e^+e^- \to \pi^+\pi^-}},
\label{defr}
\eeqn
where $\sigma^{I=1}_{\rm e^+e^-}$ is the total cross section of $\ee$ 
annihilation into hadrons with isospin 1, 
$\sigma_{\rm e^+e^- \to \pi^+\pi^-}$ and
$\sigma^{\rm I=1}_{\rm  non-2\pi}$ are
the cross section of the process $\ee \to \pi^+\pi^-$ and
that of $\ee$ annihilation into hadronic states with I=1 other than
$\pi^+\pi^-$, respectively. 
In the energy region $\sqrt{s} < 4m_{\pi}$, where
$\sqrt{s}$ is the $\ee$ c.m. energy,
 $\eta_1 \equiv 1$ and the 
inequality above just reduces to the Watson theorem: $\psi=\delta_1$.
At higher energies, when $\sqrt{s} > 4m_{\pi}$, it is a generalization 
of the Watson theorem, so that at any fixed energy  
a relation constraining $\eta_1$ and $|\psi~-~\delta_1|$ can be obtained
from Eq.~(\ref{gen1}) using the corresponding value of $r$.
Once some reliable analysis of $\pi\pi$ scattering yields
$\eta_1$ and $\delta_1$, the inequality~(\ref{gen1}) provides us with
bounds on possible values of the pion form factor phase $\psi$, such 
bounds may be of importance for theoretical models of the form 
factor~\cite{ph1,ph2,ph3,ph4,ph5} and may serve as an estimate of the 
feasibility of future interference type measurements of the phase.
These bounds will be discussed further in this letter and now we will 
discuss constraints 
imposed by $\ee$ experiments only.
Let us notice that  
the following consequences of the relation~(\ref{gen1}) can
be derived at $r < 1$: 
\beqn
\eta_1 \geqslant \frac {(1-r)} {(1+r)},
\label{in1}
\eeqn
and for the phase difference
\beqn
sin^2(\psi-\delta_1) \leqslant \frac {1} {2} [1-\sqrt{1-r^2}].
\label{in2}
\eeqn
In 1973, when these bounds appeared, studies of $\ee$ annihilation into 
hadrons only started. Now, thirty years later, we can quantify 
these bounds using experimental information accumulated since 
that time, particularly recent data on the I=1 final states 
from the VEPP-2M collider in Novosibirsk~\cite{review}. 

For the $2\pi$ channel there is a new high precision measurement
between 0.61 and 0.96 GeV from CMD-2 in which the pion form factor 
was determined with a systematic uncertainty of 0.6\%~\cite{cmd2pi}. 
At higher energies one can use the older data set from OLYA~\cite{olya}. 
Channels with non-2$\pi$ hadronic production below 1 GeV or more 
precisely, below the threshold of kaon pair production, 
$\sqrt{s} \leqslant 2m_{K}$, are saturated by $\ee$ annihilation into 
the $2\pi^+2\pi^-, \pi^+\pi^-2\pi^0$ and $\omega\pi$ final states.
For the two former there exist CMD-2
measurements~\cite{cmd4pi1,cmdpak} whereas for the latter 
independent results from 
CMD-2~\cite{cmdpak,cmdomp} and SND~\cite{sndomp} are 
available. Since the dominant decay mode
$\omega \to \pi^+\pi^-\pi^0$ is already taken into account in the 
cross section of the process $e^+e^- \to \pi^+\pi^-2\pi^0$, to avoid
double counting one should take only a part of the cross section
of the process $e^+e^- \to \omega\pi^0$, namely the one corresponding 
to the $\omega \to \pi^0\gamma$ and $\omega \to \pi^+\pi^-$ decays with
the branching ratios from~\cite{pdg}.

\begin{figure}
\begin{center}
\includegraphics[width=0.6\textwidth]{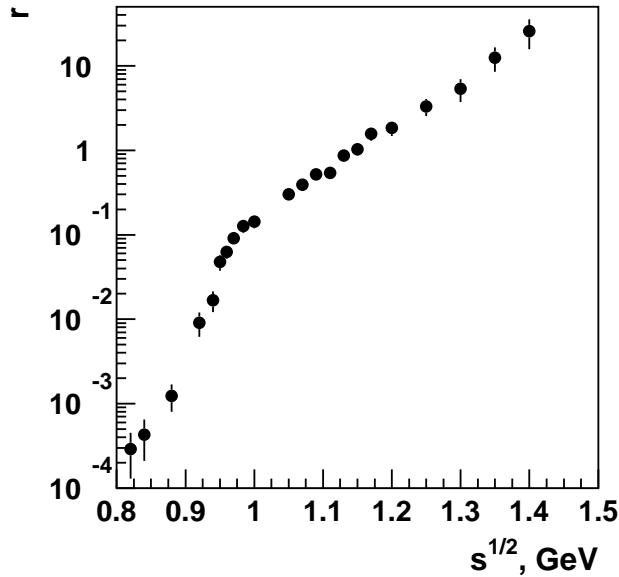}
  \caption{Energy dependence of $r$, the ratio of the non-2$\pi$ 
cross section of $e^+e^- \to$ hadrons with isospin 1 to the 
cross section of the process $e^+e^- \to \pi^+\pi^-$, see its
definition in Eq.~(\ref{defr}).}
  \label{figr}
\end{center}
\end{figure}

The situation is more complicated at energies above the threshold 
of kaon pair production. The measured cross sections of 
$\ee \to K_S K_L$~\cite{kskl} and $\ee \to K^+K^-$~\cite{kch} 
contain both isovector and isoscalar components
that are usually separated within some specific model only. Since we are
mostly interested in the energy range where $r < $ 1, or 
$\sqrt{s} \lesssim 1.15$ GeV, one can assume that the isoscalar 
component dominates due to the closeness of the $\phi$ meson. 
This assumption is confirmed by a simple SU(3) based estimate of 
the I=1 part of the kaon cross section~\cite{su3}:
\beqn
\sigma^{I=1}_{K\bar{K}}(s)=\sigma_{\pi\pi}(s)
\frac {\beta^3_{K^0}~+~\beta^3_{K^-}} {4\beta^3_{\pi^-}},
\eeqn
where 
\beqn
\beta=\sqrt{1~-~4m_X^2/s},~~~~ X=\pi^-,K^0,K^-.
\eeqn
The estimate shows that this cross section is numerically less than 5\% 
of the total cross section with I=1 at the highest energy considered and 
negligible close to the threshold. For safety, it is added  
to $\sigma^{\rm I=1}_{\rm  non-2\pi}$. For two four-pion and 
$\omega\pi$ channels above 1 GeV we use the data from the 
CMD-2~\cite{cmd4pi2,cmdomp} and SND~\cite{snd4pi,sndomp}. 
Above 1.25 GeV one should also add the contribution of the 
$\eta\pi^+\pi^-$ final state using the results of ND~\cite{nd}
and CMD-2~\cite{cmdeta}. The final states with six pions are known 
to have small cross sections below 1.4 GeV~\cite{barkov}.
Summing the contributions from different 
channels mentioned above and averaging results from different groups
in the overlapping energy ranges one finally
obtains the quantity $r$ as a function of energy shown in 
Table~\ref{tabr} and Fig.~\ref{figr}. For convenience we also separately
present in Table~\ref{tabr} the 2$\pi$ and non-$2\pi$ parts of the
cross section. All errors correspond to  
the statistical and systematic uncertainties combined in quadrature. 
It can be seen that the non-2$\pi$ cross section grows fast and 
at the c.m. energy $\sqrt{s} \approx$ 1.15 GeV becomes 
larger than the 2$\pi$ one so that $r > 1$ and the 
inequalities~(\ref{in1}) and (\ref{in2}) no 
longer contain non-trivial information. The accuracy with which 
we currently know the quantity $r$, is limited at 
$\sqrt{s} \lesssim 0.95$~GeV by poor precision of the non-2$\pi$
cross section, whereas it is the 2$\pi$ cross section which 
limits the accuracy at $\sqrt{s} > 1.1$~GeV since at the moment 
it is based on the old measurement with the OLYA
detector~\cite{barkov}. We do not show in Table ~\ref{tabr} energies 
above 1.4~GeV where data suffer from large
uncertainties~\cite{review}. As we will see, this energy region
is not practically interesting since at large $r$ the inequalities
discussed are not constraining the inelasticity and phase difference
any longer.  
  
\begin{table*}
\begin{center}
\caption{$\ee$ cross sections $\sigma_{e^+e^- \to \pi^+\pi^-}$,
$\sigma^{\rm I=1}_{\rm non-2\pi}$ and the ratio $r$ 
 as a function of energy}
\medskip
\begin{tabular}{cccc}
\hline
$\sqrt{s}$, GeV & $\sigma_{e^+e^- \to \pi^+\pi^-}$, nb & 
$\sigma^{\rm I=1}_{\rm non-2\pi}$, nb & $r$ \\
\hline
0.82 & 619.0 $\pm$ 20.3 & 0.18 $\pm$ 0.10 & $(2.9\pm1.6) \cdot 10^{-4}$ \\
\hline
0.84 & 415.4 $\pm$ 17.5 & 0.18 $\pm$ 0.09 & $(4.3\pm2.2) \cdot 10^{-4}$ \\
\hline
0.88 & 233.6 $\pm$ 18.2 & 0.29 $\pm$ 0.10 & 
$(1.24\pm0.44)\cdot 10^{-3}$ \\
\hline
0.92 & 129.7 $\pm$ 6.5 & 1.18 $\pm$ 0.37 & $(0.91\pm0.29)\cdot 10^{-2}$ \\
\hline
0.94 & 106.8 $\pm$ 5.1 & 1.78 $\pm$ 0.48 & $(1.67\pm0.46)\cdot 10^{-2}$ \\
\hline
0.95 & 92.9 $\pm$ 4.8 & 4.43 $\pm$ 0.92 & $(4.77\pm1.02)\cdot 10^{-2}$ \\
\hline
0.96 & 86.2 $\pm$ 4.6 & 5.42 $\pm$ 0.93 & $(6.29\pm1.13)\cdot 10^{-2}$ \\
\hline
0.97 & 76.3 $\pm$ 7.4 & 6.91 $\pm$ 1.16 & $(9.06\pm1.40) \cdot 10^{-2}$ \\
\hline
0.984 & 63.0 $\pm$ 6.5 & 8.00 $\pm$ 1.14 & $0.127\pm0.022$ \\
\hline
1.00 & 60.3 $\pm$ 5.8 & 8.62 $\pm$ 1.16 & 0.143 $\pm$ 0.024 \\
\hline
1.05 & 43.3 $\pm$ 4.3 & 13.05 $\pm$ 1.23 & 0.30 $\pm$ 0.04 \\
\hline
1.07 & 37.1 $\pm$ 4.6 & 14.49 $\pm$ 1.30 & 0.39 $\pm$ 0.06 \\
\hline
1.09 & 33.6 $\pm$ 3.9 & 17.41 $\pm$ 1.46 & 0.52 $\pm$ 0.07 \\
\hline
1.11 & 32.4 $\pm$ 3.9 & 17.60 $\pm$ 1.38 & 0.54 $\pm$ 0.08 \\
\hline
1.13 & 23.2 $\pm$ 3.1 & 20.04 $\pm$ 1.54 & 0.86 $\pm$ 0.13 \\
\hline
1.15 & 22.0 $\pm$ 3.2 & 22.65 $\pm$ 1.74 & 1.03 $\pm$ 0.17 \\
\hline
1.17 & 16.3 $\pm$ 2.6 & 25.52 $\pm$ 1.76 & 1.57 $\pm$ 0.27 \\
\hline
1.20 & 16.1 $\pm$ 3.0 & 29.66 $\pm$ 1.95 & 1.84 $\pm$ 0.36 \\
\hline
1.25 & 11.0 $\pm$ 2.4 & 36.43 $\pm$ 2.04 & 3.31 $\pm$ 0.75 \\
\hline
1.30 & 8.1 $\pm$ 2.2 & 43.49 $\pm$ 2.47 & 5.37 $\pm$ 1.62 \\
\hline
1.35 & 3.8 $\pm$ 1.2 & 47.71 $\pm$ 3.03 & 12.56 $\pm$ 4.04 \\
\hline
1.40 & 2.1 $\pm$ 0.8 & 54.03 $\pm$ 3.33 & 25.73 $\pm$ 9.93 \\
\hline
\end{tabular}
\label{tabr}
\end{center}
\end{table*}     
  
\begin{figure}
\begin{center} 
\hspace*{5mm}
\begin{minipage}{16.cm}
\begin{tabular}{c@{\hspace*{5mm}}c}
\psfig{figure=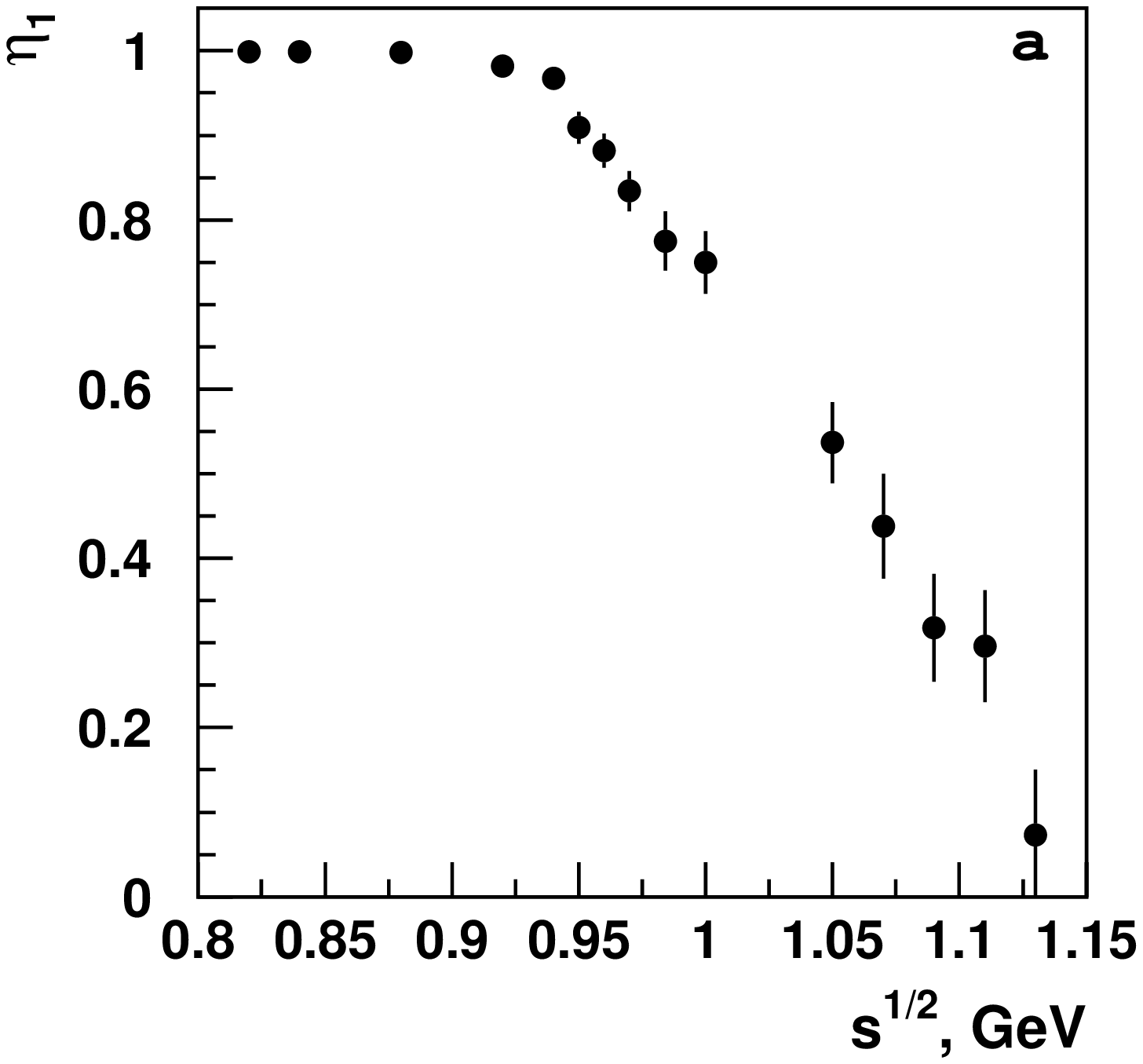,width=65mm}
&
\psfig{figure=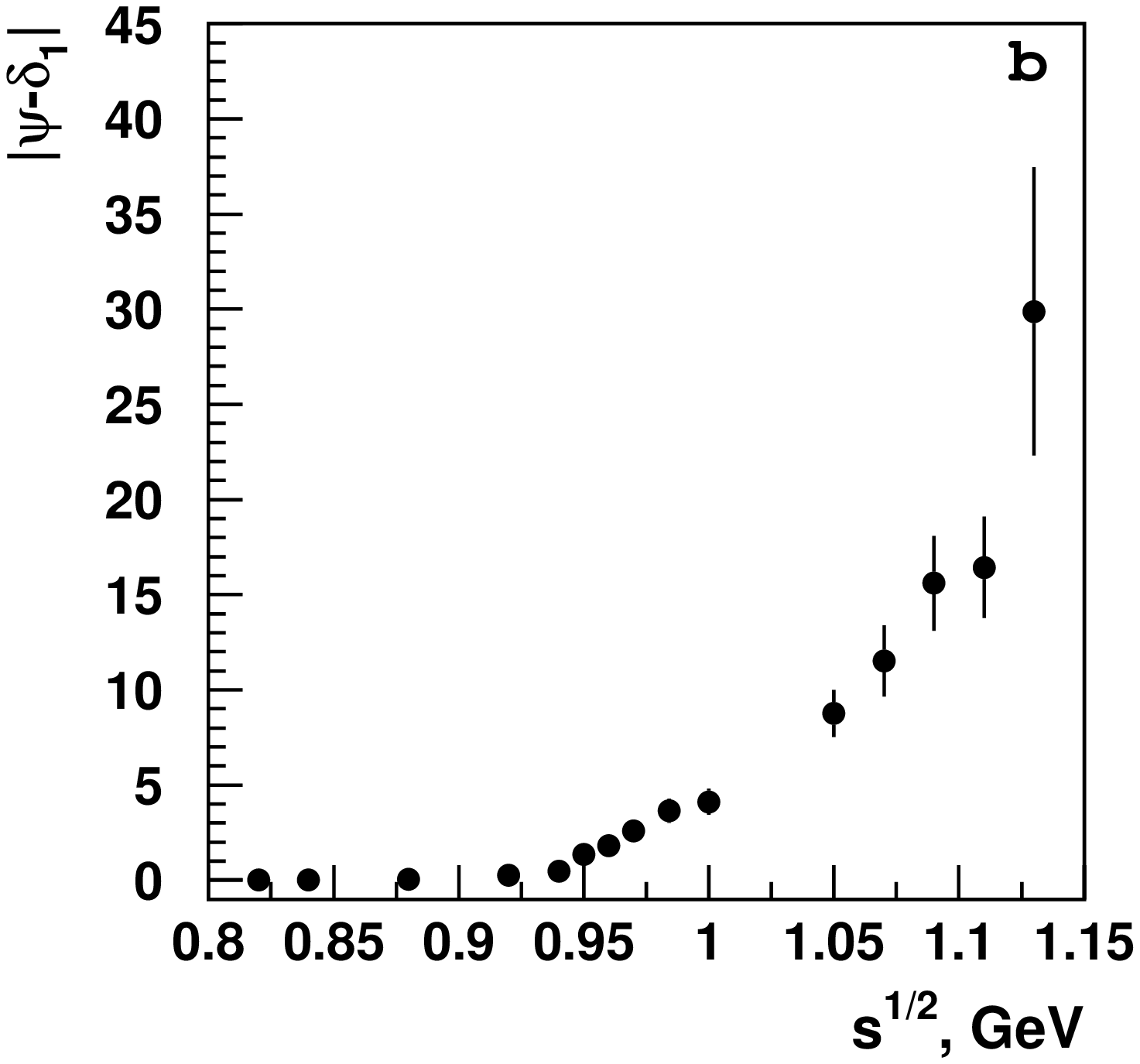,width=65mm}
\end{tabular}
\end{minipage}
\caption{Energy dependence of: a/ the elasticity $\eta_1$. The
allowed area is above the points with errors; b/ the absolute
value of the difference of the
form factor phase $\psi$ and the phase shift $\delta$ in degrees.
The allowed area is below the points with errors. The phase difference
can be determined up to $\pm k\pi$, where $k$ is integer.}    
\label{fig2}
\end{center}
\end{figure}

In Figs.~\ref{fig2}a and~\ref{fig2}b  we present the results of applying 
the inequalities (\ref{in1}) and (\ref{in2}), respectively. In 
Fig.~\ref{fig2}a the
allowed range for the elasticity extends from the lower limit
given by the points with errors recalculated from~(\ref{in1}) to 
its maximum possible value equal to 1.
It can be seen that this range is rather narrow and provides strict
constraints at low energies when the process 
$e^+e^- \to \pi^+\pi^-$ dominates the isospin 1 cross section of
$\ee$ annihilation. With the fast growth of the non-2$\pi$ cross
section above 1 GeV the constraints provided by the relation~(\ref{in1})
become rather loose.  

In Fig.~\ref{fig2}b the
allowed range for the phase difference extends from $0^{\circ}$ to
its maximum possible value given by points recalculated 
from~(\ref{in2}). 
Similarly to the previous case, strict bounds are obtained at low
energies. It can be seen that up to $\sim$ 0.95 GeV the phase
difference is close to zero. When energy grows, the maximum 
phase difference also becomes larger and tends to 45$^{\circ}$ 
at $r=1$ or $\sqrt{s} \approx$ 1.15 GeV. Note that the maximum value
of the phase difference derived from the inequality~(\ref{in2}) 
can be determined up to $\pm k\pi$, where $k$ is integer.  
  

\begin{figure}
\begin{center} 
\hspace*{5mm}
\begin{minipage}{16.cm}
\begin{tabular}{c@{\hspace*{5mm}}c}
\psfig{figure=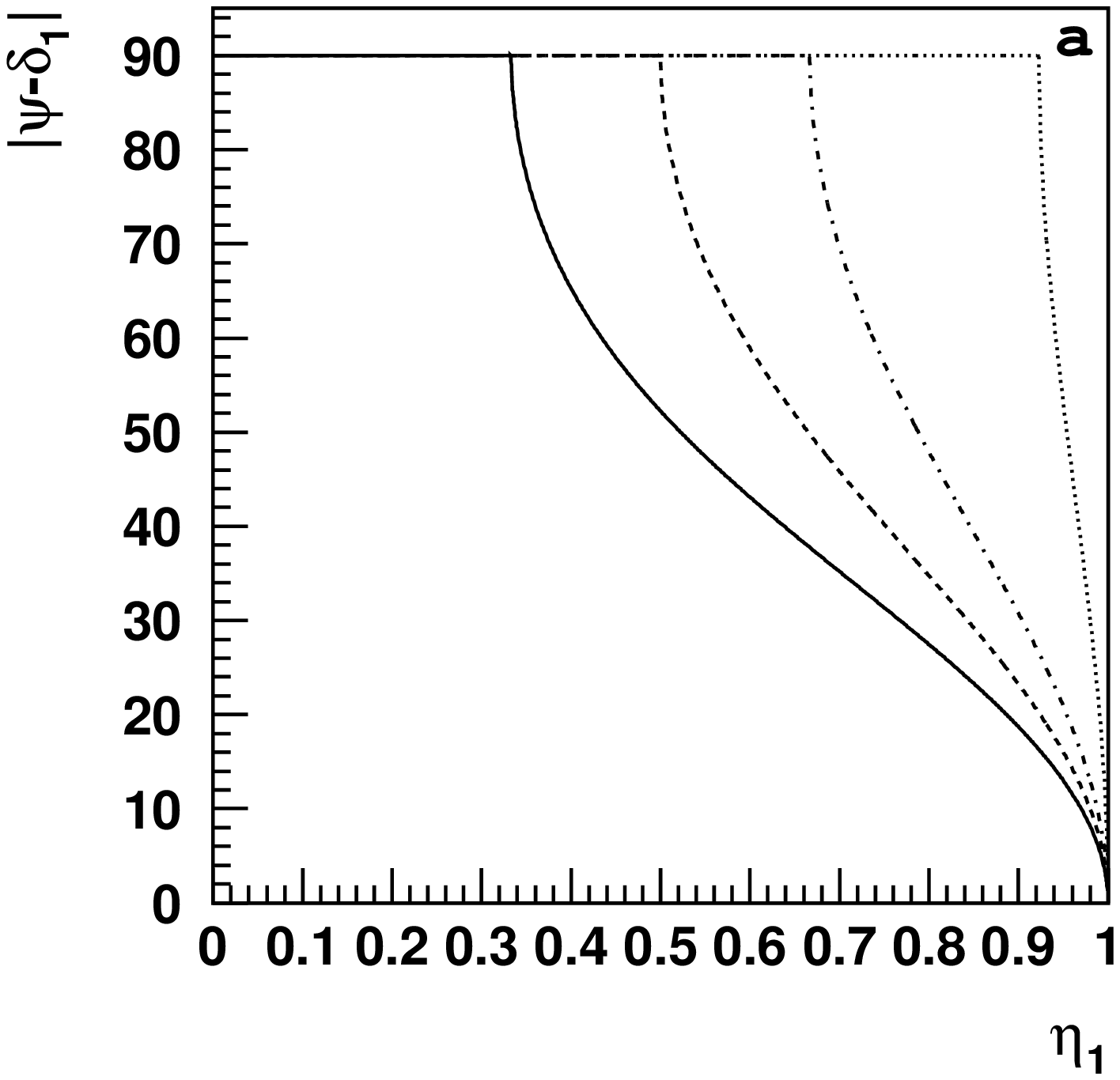,width=65mm}
&
\psfig{figure=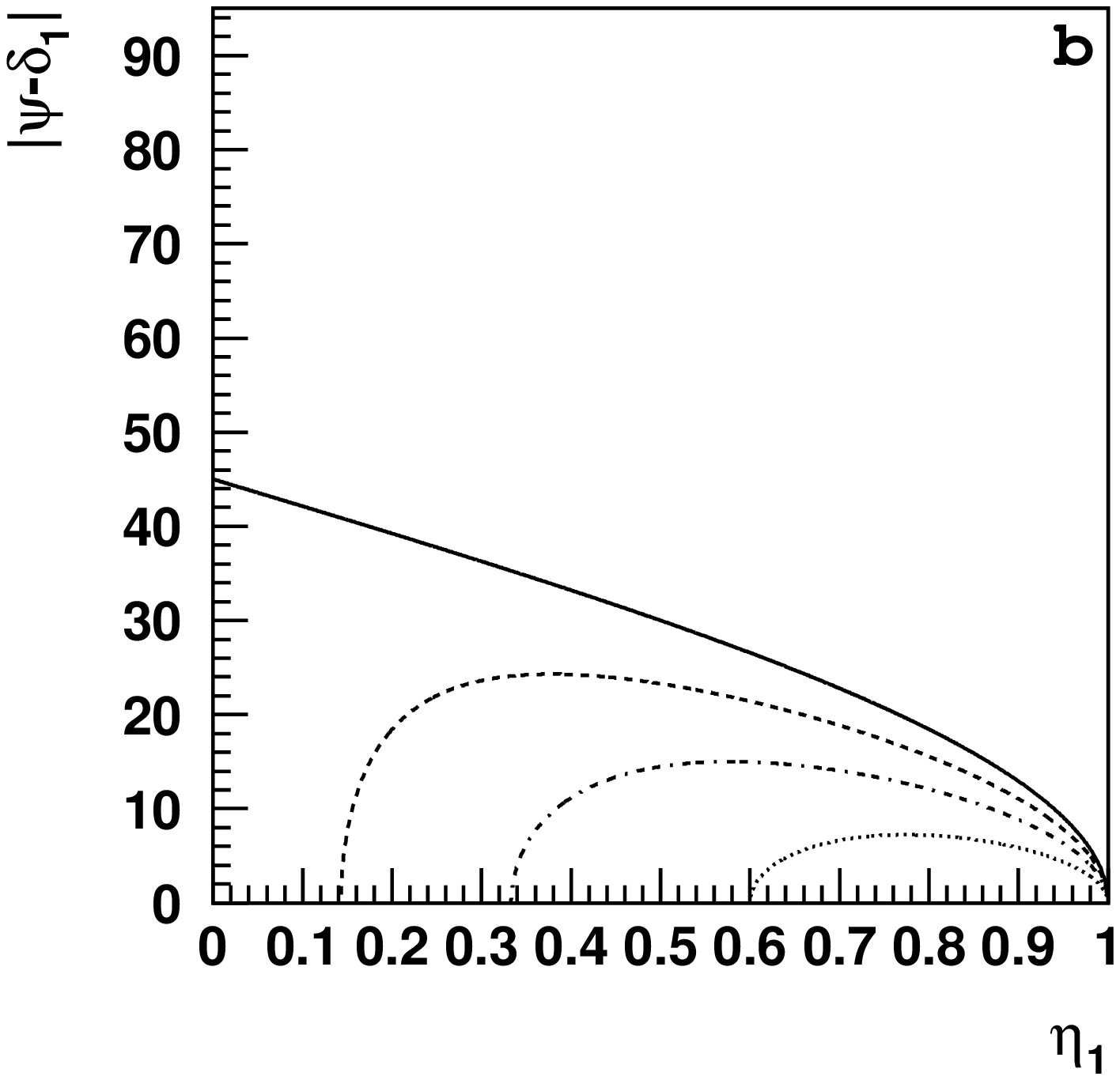,width=65mm}
\end{tabular}
\end{minipage}
\caption{The absolute value of the phase difference  $|\psi~-~\delta_1|$ 
versus the elasticity 
$\eta_1$ for different values of $r$: a/ $r > 1$, 
the solid, dashed, dashed-dotted and dotted curves correspond to
$r = 2, 3, 5$ and 25, respectively; 
b/ $r \leqslant 1$.
the solid, dashed, dashed-dotted and dotted curves correspond to
$r = 1, 0.75, 0.5$ and 0.25, respectively. 
In all cases the allowed area is below the curve.}
\label{fig3}
\end{center}
\end{figure}

The general relation (\ref{gen1}) bounding 
both $\eta_1$ and $|\psi~-~\delta_1|$ holds at any positive $r$. 
Figure~\ref{fig3}
illustrates that non-trivial constraints are imposed by (\ref{gen1})
for the whole range of  $r$  values from 
Table~\ref{tabr}.
For each fixed $r > 1$ (Fig.~\ref{fig3}a) the following pattern is 
observed: below certain value of the elasticity $\eta^c_1$ 
the phase difference can vary from 0$^{\circ}$ to its maximum of
90$^{\circ}$. At larger $\eta_1$ the maximum allowed value of the
phase difference starts decreasing and falls down to $0^{\circ}$ 
at $\eta_1=1$. The larger is $r$, the larger is the $\eta^c_1$.
At $r \sim 25$ the value of $\eta^c_1$ is very close to 1, i.e.
the whole plane of   $\eta_1$ and $|\psi~-~\delta_1|$   is allowed.
Because of that we do not consider the energy range above 1.4 GeV
where the value of $r$ is even larger.   
The behaviour is different at $r \leqslant 1$. At $r=1$ the curve of
maximum phase difference goes down monotonously from 45$^{\circ}$ 
at $\eta_1 \rightarrow 0$ to $0^{\circ}$ at $\eta_1=1$.
At any $r < 1$ it starts from  $0^{\circ}$ at some minimal $\eta_1$, 
grows with $\eta_1$, reaches its maximum ($ < 45^{\circ}$) at some 
$\eta_1$ and then falls to zero at $\eta_1=1$.
As above, the area under the curve is allowed.

It is tempting to check the efficiency of the inequality~(\ref{gen1})
by using additionally the results of the analysis of
the $\pi\pi$ scattering by the CERN-Munich group~\cite{hy1,hy2}.
By simply rewriting~(\ref{gen1}), one can obtain an upper bound for
$|\psi - \delta_1|$ as a function of $r$ and $\eta_1$. Using the
results of the CERN-Munich group for the elasticity~\cite{hy1,hy2} 
and our results for $r$ from Table~\ref{tabr}, we calculate the bound
as shown in Table~\ref{tabb}. Now, when the bound is determined,
one can solve the inequality~(\ref{gen1}) and obtain the allowed
range for the form factor phase $\psi$  also presented in 
Table~\ref{tabb}. The width
of this range was enlarged in both directions by one standard deviation
calculated by combining the uncertainty of the bound above and 
that of $\delta_1$ from Refs.~\cite{hy1,hy2} in quadrature. 

\begin{table*}
\begin{center}
\caption{Bound on the absolute value of the phase difference and the
allowed range for the pion form factor phase  as a 
function of energy}
\begin{tabular}{ccc}
\hline
$\sqrt{s}$, GeV & $|\psi - \delta_1|_{\rm {max}}$ & $\psi$ \\ 
\hline       
0.95   &   1.2  $\pm$         0.9   &  146$^\circ$--152$^\circ$   \\
\hline 
0.97   &   2.4  $\pm$          2.2  &  144$^\circ$--160$^\circ$  \\
\hline 
0.99    &  3.7   $\pm$         0.5  &  150$^\circ$--160$^\circ$  \\
\hline 
1.02    &  5.9  $\pm$          0.7  &   153$^\circ$--167$^\circ$  \\
\hline 
1.06    &  8.4  $\pm$          0.7  &  154$^\circ$--173$^\circ$  \\
\hline 
1.10   &  10.3   $\pm$         0.7  &  153$^\circ$--176$^\circ$   \\
\hline 
1.14  & 14.9  $\pm$         1.1    &  152$^\circ$--185$^\circ$ \\
\hline 
1.18    & 15.1   $\pm$         1.4  &  149$^\circ$--182$^\circ$   \\
\hline 
1.22   &  20.8   $\pm$         2.0  &  147$^\circ$--192$^\circ$   \\
\hline 
1.26   &  33.0   $\pm$         3.7  &  136$^\circ$--209$^\circ$   \\
\hline 
1.30    & 33.8   $\pm$         6.1  &  131$^\circ$--211$^\circ$  \\
\hline 
1.34   &  79.5   $\pm$        46.7  &   43$^\circ$--299$^\circ$   \\
\hline 
1.38  &   52.7   $\pm$        13.1  &  106$^\circ$--235$^\circ$   \\
\hline 
\end{tabular}
\label{tabb}
\end{center}
\end{table*}

\begin{figure}
\begin{center}
  \includegraphics[width=0.6\textwidth]{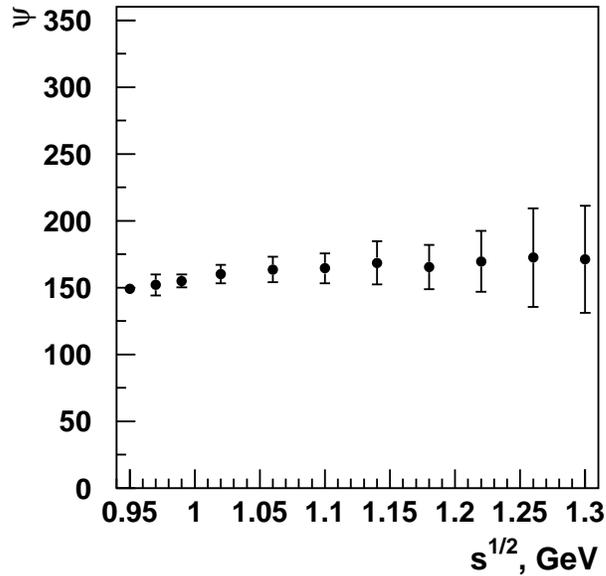}
\caption{Pion form factor phase as a function of energy.
The allowed range of the phase is shown by points with error bars,  
so that an error bar equals half a width of the range.  }
\label{phase}
\end{center}
\end{figure}

One can see that at low enough energies the bound following from
the inequality~(\ref{gen1}) is really robust and allows a
meaningful estimate of the pion form factor phase. This is true
until approximately 1.2 GeV. At higher energies the central value
of the bound  as well as its error become too large and no longer
provide a good estimate of the phase $\psi$. In Fig.~\ref{phase}
we show the allowed range for the phase
as a function of energy.

The inequalities discussed above were originally obtained 
for a general case
of any reactions involving a weakly interacting channel; see, e.g.,
their application to the case of  
$\gamma\gamma \to \pi\pi$ scattering~\cite{lyth}. Similar relations
were obtained in a coupled-channel approach in Ref.~\cite{bas}.

In conclusion, we would like to emphasize that the bounds using
information on $\ee \to$ hadrons alone  already  yield quite
stringent limitations exhibited in Figs.~\ref{fig2}a,b.
In particular, the solution A of Ref.~\cite{bugg} for the
$\pi\pi$ elasticity $\eta_1$ yielding the value of $1-\eta_1$ at the 
level of a few \% in the vicinity of the $\rho$ meson resonance 
can be ruled out by $\ee$ data, see Fig.~\ref{fig2}a.
There is also a permanent question how far the Watson equality 
($\psi=\delta_1$) is valid, 
see, e.g., Ref.~\cite{roy,ph1,ph2,ph3,ph4,ph5}. Our Fig.~\ref{fig2}b
gives bounds on deviations from the Watson equality for energies
well above 4$m_{\pi}$.  

One can hopefully further elaborate the obtained bounds 
after the analysis of high statistics data samples in Novosibirsk is 
completed~\cite{review}.
  
The authors would like to express their gratitude to the University of
Siegen, Germany where they met and discussed for the first time the 
possibility of this work. They are indebted to P.P.~Krokovny for
his help while preparing a manuscript.
L.{\L}. was supported in part by the European 
Community's Human Potential Programme under contract HPRN-CT-2002-00311
EURIDICE. S.E. was supported in part by the grant RFBR-03-02-16843. 
S.E. is grateful to the Institute for Nuclear Studies, \'{S}wierk, 
Poland for its hospitality during the time when this work was completed.

\end{document}